\documentclass[prl,10pt,twocolumn,preprintnumbers,amsmath,amssymb,superscriptaddress,noshowpacs]{revtex4}
\usepackage{amsmath}
\usepackage{amssymb}
\usepackage{amsfonts}
\usepackage{eucal}
\usepackage[active]{srcltx}%
\usepackage{color}
\usepackage[all]{xy}
\usepackage{epsfig,graphicx,bm}
\usepackage[colorlinks=true, linkcolor=blue,urlcolor=magenta, filecolor=green,
      citecolor=red,
      pdfstartview=FitV,
      pdftitle={},
        pdfauthor={Hamid Afshar, Daniel Grumiller, Shahin Sheikh-Jabbari},
        pdfsubject={},
        pdfkeywords={},
        pdfpagemode=None,
        bookmarksopen=true
      ]{hyperref}
\usepackage{color}
\usepackage[normalem]{ulem}

\newcommand{\cT}{\mathcal T}
\newcommand{\cY}{\mathcal Y}
\newcommand{\cJ}{\mathcal J}
\newcommand{\cK}{\mathcal K}
\newcommand{\cB}{\mathcal B}

\newcommand{\cL}{{\mathcal{L}}}

\newcommand{\<}{\langle}
\renewcommand{\>}{\rangle}

\newcommand{\De}{\Delta}

\newcommand{\eq}[2]{\begin{equation} #1 \label{#2} \end{equation}}
\newcommand{\ba}{\begin{array}}
\newcommand{\ea}{\end{array}}
\newcommand{\be}{\begin{equation}}
\newcommand{\ee}{\end{equation}}
\newcommand{\bea}{\begin{eqnarray}}
\newcommand{\eea}{\end{eqnarray}}
\newcommand{\bse}{\begin{subequations}}
\newcommand{\ese}{\end{subequations}}
\newcommand{\bi}{\begin{itemize}}
\newcommand{\ei}{\end{itemize}}

\definecolor{darkgreen}{rgb}{0,0.3,0}
\definecolor{darkblue}{rgb}{0,0,0.3}
\definecolor{darkred}{rgb}{0.7,0,0}

\newcommand{\old}[1]{}

\begin{document}



\title{Black Hole Horizon Fluff:
\\ Near Horizon Soft Hairs as  Microstates of Three Dimensional Black Holes}

\author{H.~Afshar}
\email{afshar@ipm.ir}
\affiliation{School of Physics, Institute for Research in Fundamental Sciences (IPM),  P.O.Box 19395-5531, Tehran, Iran}

\author{D.~Grumiller}
\email{grumil@hep.itp.tuwien.ac.at}
\affiliation{Institute for Theoretical Physics, TU Wien, Wiedner Hauptstrasse 8--10/136, A-1040 Vienna, Austria}

\author{M.M.~Sheikh-Jabbari}
\email{jabbari@theory.ipm.ac.ir}
\affiliation{School of Physics, Institute for Research in Fundamental Sciences (IPM),  P.O.Box 19395-5531, Tehran, Iran}


\begin{abstract}

We provide the first explicit proposal for all microstates of generic black holes in three dimensions (of Ba\~nados--Teitelboim--Zanelli-type): black hole microstates, termed `horizon fluff', are a particular class of near horizon soft hairs which have zero energy as measured by the horizon observer and cannot be distinguished by observers at finite distance from the horizon. 
These states are arranged in orbits of the two-dimensional conformal algebra associated with the asymptotic black hole
geometry. We count these microstates using the Hardy--Ramanujan formula for the number of partitions of a given integer into non-negative integers,  recovering the Bekenstein--Hawking entropy. We discuss possible extensions of our black hole microstate  construction to astrophysical Kerr-type black holes.

\end{abstract}


\maketitle

\enlargethispage{0.5truecm}

\paragraph{\textbf{Introduction.}} Classic works of early 1970's uncovered a less black side of the black holes (BHs): BHs radiate, which to a very good extent is black body radiation at a given temperature \cite{Hawking:1974sw}, they have  entropy $S_{\textrm{\tiny BH}}$ proportional to the area $A$ of the event horizon \cite{Bekenstein:1973ur} ($G_N$ is Newton's constant)
\eq{
S_{\textrm{\tiny BH}} = A/(4G_N)
}{eq:SBH}
and follow laws of thermodynamics \cite{Bardeen:1973gs}. The thermodynamic behavior is ubiquitous for BH solutions to diffeomorphism invariant gravity theories \cite{Wald:1999vt}. 

Hawking radiation generically leads to evaporation of BHs and this brings about the information paradox: the process of formation, radiation and evaporation of BHs apparently is not unitary \cite{Mathur:2008wi}. Non-unitarity of the BH dynamics may be resolved and understood in the same way it is explained in any thermodynamical system, namely as an artifact of the thermodynamic limit. For this idea to work one needs to identify the underlying statistical mechanical system: the BH microstates that account for the macroscopic entropy \eqref{eq:SBH}. 

There are uniqueness and no-hair theorems for BH solutions in Einstein gravity: for a given set of charges like mass, angular momentum and electric (or magnetic) {conserved} charges and specified asymptotic behavior there is a unique solution to the theory \cite{Mazur:2000pn}. This  suggests that BH microstates, if they exist, may not be found within the set of solutions to the Einstein theory.  {Therefore}, in search for BH microstates, the main focus has been on quantum gravity theories. Among them the most successful one has been string theory, where a BH is modeled through a combination of strings and branes that carry the same classical charges \cite{Strominger:1996sh}. See e.g.~\cite{Sen:2007qy} for more on BH microstate counting and string theory. However, the cases where all BH microstates can be identified explicitly are rare. A less explicit but more universal approach pointed out by Strominger \cite{Strominger:1997eq} and  mainly advocated by Carlip (e.g. see  \cite{Carlip:1994gy,Carlip:1998wz}) is inspired by Cardy's work on two-dimensional (2D) conformal field theories (CFTs) \cite{Cardy:1986ie, Bloete:1986qm} and exploits symmetries to perform microstate counting without really identifying  them.

It is thus fair to say that the problem of explicitly constructing BH microstates remains largely unsolved, particularly for BHs at finite temperature. In the present Letter we pursue and implement recent ideas \cite{Hawking:2016msc, Sheikh-Jabbari:2016lzm} to explicitly construct for the first time all microstates of a specific family of BHs at finite temperature. While some details of our construction are specific to this family, the overall setup could work in generality. We shall provide some evidence that this is indeed the case.

One key ingredient is the ``soft hair'' proposal by Hawking, Perry and Strominger \cite{Hawking:2016msc}, whose work has engendered a lot of research activities, see for instance \cite{Penna:2015gza,  Afshar:2015wjm, Hooft:2016itl, Bianchi:2016tju, Averin:2016ybl, Compere:2016jwb, Dai:2016sls,  Compere:2016hzt, Afshar:2016wfy, Sheikh-Jabbari:2016unm, Kapec:2016aqd, Sheikh-Jabbari:2016lzm,  Eling:2016xlx,   Giddings:2016plq,  Compere:2016gwf, Hotta:2016qtv, Mao:2016pwq, Setare:2016vhy, Setare:2016jba, Casadio:2016zpl, Averin:2016hhm}. This proposal suggests that BH microstates could be related to ``soft hair'', i.e., zero energy excitations on the horizon. The obvious problem with this idea is that without a cutoff on the soft hair spectrum there will be infinitely many such excitations, leading to an infinite entropy, thereby contradicting \eqref{eq:SBH}. In the present Letter we solve this problem by introducing a cutoff on the spectrum in a controlled way through a comparison between near horizon and asymptotic observables. 

The other key ingredients are the near horizon boundary conditions proposed in \cite{Afshar:2016wfy}, which lead to surprisingly simple near horizon symmetries, namely infinite copies of the Heisenberg algebra (see \cite{Donnay:2015abr} for an alternative proposal). We exploit this algebra to generate descendants of physical states that we then interpret as BH microstates. Moreover, as we shall demonstrate, this algebra, together with asymptotic information, naturally suggests a specific cutoff on the soft hair spectrum. This then establishes our main goal, namely an explicit construction of all microstates of the BH family that we consider, which is the main result of the present Letter.

The final ingredient is purely combinatorial and serves as a cross check that the number of our BH microstates matches with \eqref{eq:SBH} in the semi-classical limit. Here we use well-known mathematics results by Hardy and Ramanujan, whose relevance for BH microstate counting was already stressed a long time ago by Carlip, see e.g.~\cite{Carlip:1998qw} and Refs.~therein.

\paragraph{\textbf{Near horizon algebra and Hilbert space.}}
We focus for now on Einstein gravity in three spacetime dimensions (3D) with negative cosmological constant, with the intention of constructing all microstates of non-extremal Ba\~nados--Teitelboim--Zanelli (BTZ) BHs \cite{BTZ}. 

While everything we do has a geometric interpretation, it is simpler to work on the field theory side. By this we mean that rather than writing down metrics and discussing geometric properties we work directly with the near horizon symmetry algebra established in \cite{Afshar:2016wfy}~\footnote{In \cite{Afshar:2016wfy} the generators $\cJ_n^\pm$ has a different normalization. Our current normalization 
is the more appropriate one as it eliminates the dependence on the AdS radius, which is expected not to play an important role in the near horizon theory.}.
\be\label{NH-algebra-J}
[\cJ_n^\pm,\,\cJ_m^\pm]=\tfrac12 \,n\, \delta_{n,-m}
\ee

To construct the ``near horizon Hilbert space'' we start with the near horizon vacuum state $|0\rangle$. It is natural to define it as highest weight state with vanishing eigenvalues of $\cJ_0^\pm$, i.e., 
 $\cJ_n^\pm|0\rangle = 0$ for all $n\geq 0$.

Equation \eqref{NH-algebra-J} is the algebra of creation-annihilation operators for a free 2D boson theory on $\mathbb{R}\times S^1$ which is a CFT$_2$. The Fourier modes of its energy-momentum tensor are  given as
\be\label{NH-Vir-geb}
\cL_n^\pm\equiv\sum_{p\in\mathbb{Z}}\colon\!\cJ_{n-p}^\pm\,\cJ_p^\pm\colon
\ee 
where $\colon\colon$ denotes normal ordering. It is well-known that $\cL_n^\pm$ form Virasoro algebras of central charge one:
\begin{subequations}
\label{NH-algebra-Vir}
\begin{align}
[\cL_n^\pm,\cL_m^\pm]&=(n-m)\cL_{n+m}^\pm+\tfrac{1}{12}(n^3-n)\delta_{n,-m},\\
 [\cL_n^\pm,\cJ_m^\pm]&=-m \cJ_{n+m}^\pm.
  \end{align}
\end{subequations}
Given our vacuum definition $\cJ_n^\pm|0\rangle = 0$ for all $n\geq 0$, we deduce that the vacuum has zero eigenvalues $\cL_0^\pm |0\rangle = 0$.

\paragraph{\textbf{Descendant soft hairs.}} 
With the above we then define a generic descendant of the vacuum,  $|\Psi(\{n_i^\pm\})\rangle$, using creation operators  $\cJ_{-n_i^\pm}^\pm$ with sets of positive integers $\{n^\pm_i>0\}$, i.e. 
\be\label{generic-NH-state}
|\Psi(\{n_i^\pm\})\rangle = \!\!\prod_{\{n_i^\pm>0\}}\!\!\big( \cJ_{-n_i^+}^+\cJ_{-n_i^-}^-\big) |0\rangle\,.
\ee
All the descendants have some positive eigenvalue for $\cL_0^\pm$, which we denote by ${\cal E}_\Psi^\pm$,
\be\label{NH-energy}
\cL_0^\pm|\Psi(\{n_i^\pm)\}\rangle = \sum_i  n_i^\pm |\Psi(\{n_i^\pm\}) \rangle \equiv {\cal E}_\Psi^\pm|\Psi(\{n_i^\pm\})\rangle.
\ee

Since $\cJ_0^\pm$ commutes with all generators, the eigenvalues of $\cJ_0^\pm$ vanish for any vacuum descendant. 
Therefore,  all $|\Psi(\{n_i^\pm)\}\rangle$ have the same near horizon energy, which is measure by the Hamiltonian $H\sim \cJ_0^+ + \cJ_0^-$ \cite{Afshar:2016wfy}. This implies that all descendants \eqref{generic-NH-state} have the same energy as the vacuum, they are zero-energy excitations and, following  \cite{Hawking:2016msc}, we call them ``near horizon soft hair.'' As mentioned earlier, there are infinitely many soft hair excitations with the same energy, so without a cutoff on their spectrum any soft hair-based microstate counting will naively lead to an infinite entropy. To introduce such a cutoff we consider the asymptotic symmetry algebra of BTZ BHs and then match it with near horizon quantities.

\paragraph{\textbf{Asymptotic Virasoro algebra.}} As the seminal work of Brown and Henneaux \cite{Brown:1986nw} has revealed, the residual diffeomorphisms for asymptotically Anti-de~Sitter (AdS$_3$) geometries with their prescribed boundary conditions form two Virasoro algebras, i.e., the conformal algebra in 2D, 
\be\label{asymptotic-algebra-Vir}
[L_n^\pm,\,L_m^\pm]=(n-m)L_{n+m}^\pm+\tfrac{c}{12}\,n^3\,\delta_{n,-m} 
\ee
with $n,m\in\mathbb{Z}$ and
$c = 3\ell/(2G_N)$ 
is the Brown--Henneaux  central charge, where $\ell$ is the AdS$_3$ radius and $G_N$ is the 3D Newton constant. 

Further analysis reveals that all such locally AdS$_3$ geometries \cite{Banados:1998gg} may be labeled by their Virasoro charges (in a more technical wording, Virasoro coadjoint orbits \cite{Witten:1987ty, Balog:1997zz}) \cite{Sheikh-Jabbari:2016unm,Compere:2015knw}. 

It has been argued that this Virasoro algebra, although usually known as ``asymptotic symmetry algebra'' can in fact be recovered for generic radius away from the AdS boundary \cite{Compere:2015knw}. In what follows, however, we will conveniently call it ``asymptotic Virasoro algebra'' to distinguish it from the algebra near the horizon.

\paragraph{\textbf{Relating near horizon and asymptotic algebras.}} 
As stated in \eqref{NH-Vir-geb} there is a natural way to relate the near horizon algebra to a (near horizon) Virasoro algebra with central charge $c=1$. Now we want to related it instead to the asymptotic Virasoro algebra \eqref{asymptotic-algebra-Vir}. To this end we first restore the normalization used in \cite{Afshar:2016wfy} and rewrite the near horizon algebra \eqref{NH-algebra-J} as
\eq{
[J_n^\pm,\,J_m^\pm]=\tfrac k2 \,n\, \delta_{n,-m}
}{eq:J}
where $k=\ell/(4G_N)$.

As shown in \cite{Afshar:2016wfy} the relation to the asymptotic Virasoro algebra \eqref{asymptotic-algebra-Vir} involves a twisted Sugawara construction~\footnote{We define Fourier modes in the $-$ sector with a sign relative to \cite{Afshar:2016wfy} to have positive $\hat u(1)$ level in the $\pm$ parts of the algebra; apart from this minor change our conventions are compatible.}
\be\label{eq:twistedsugawara}
L_n^\pm\equiv inJ_n^\pm+\tfrac1{k}\sum_{p\in\mathbb{Z}} J^\pm_{n-p}J^\pm_p,
\ee
leading to the algebra \eqref{asymptotic-algebra-Vir} with \eqref{eq:J} and
\eq{
[L_n^\pm,\,J_m^\pm]=-m\,J^\pm_{n+m}+ i\tfrac k2 \,m^2\, \delta_{n,-m}.
}{eq:LJ}
Here $c=6k$ in the semi-classical (large $c$) limit. 

We have now two Virasoro algebras at our disposal, the near horizon one with unity central charge \eqref{NH-algebra-Vir} and the asymptotic one with Brown--Henneaux central charge \eqref{asymptotic-algebra-Vir}. Following Ba\~nados \cite{Banados:1998wy} it is then suggestive to relate the respective Virasoro zero modes by
\eq{
c L_0^\pm = \cL_0^\pm-\tfrac1{24}\,,\qquad L_n^\pm 
=\frac{1}{c}\cL_{nc}^\pm\,,\quad n\neq 0\,.
}{eq:LcL}
We note that with the above relations we do not map the near horizon and asymptotic currents, $\cJ_{n}^\pm$ and $J_n^\pm$, but the associated Virasoro algebras. More detailed discussions on the $L_n-\cL_n$ map may be found in \cite{Afshar:2017okz}.
This algebraic observation, as we will see, has important physical implications for BH microstates.

\paragraph{\textbf{Horizon fluff as black hole microstates.}} A non-extremal BTZ BH corresponds to a configuration with vanishing Virasoro charges $L_n^\pm$ for $n\neq 0$ and positive zero mode charges $L_0^\pm$ \cite{Sheikh-Jabbari:2016unm,Compere:2015knw}.  In terms of  the asymptotic Virasoro algebra \eqref{asymptotic-algebra-Vir} we get the vacuum expectation values (vevs)
\be\label{BTZ-L0}
\<L_0^\pm\>_{\textrm{\tiny BTZ}}= \Delta^\pm, \qquad
\< L_{n\neq 0}^\pm\>_{\textrm{\tiny BTZ}}=0 \,. 
\ee 
The vevs $\Delta^\pm$ are related to BTZ mass and angular momentum as
$\ell M_{\textrm{\tiny BTZ}}=\Delta^++\Delta^-$, $J_{\textrm{\tiny BTZ}}=\Delta^+-\Delta^-$.

The above considerations then give the most natural and surprisingly simple definition for the BTZ BH and its microstates: microstates of a BTZ BH are all states in the near horizon Hilbert space that satisfy \eqref{BTZ-L0}. Explicitly, we define the vector space of BH microstates ${\cal V}_{\cB}$ 
through the conditions
\be\label{BTZ-microstate-def}
\langle \cB'|L_{n\neq 0}^\pm |\cB\rangle=0,\ \ \forall \cB,\cB'\in {\cal V}_{\cB}
\ee
together with a normalization condition spelled out below.
Recalling the relations between asymptotic and near horizon generators \eqref{eq:LcL} one may readily observe that all solutions to \eqref{BTZ-microstate-def} are of the form
\be\label{BH-microstate}
|\cB(\{n_i^\pm\})\rangle = {\cal N}_{\{n_i^\pm\}}\!\!\!\! \prod_{\{n_i^\pm>0\}} \!\!\!\!\big(\cJ_{-n_i^+}^+ \,\cJ_{-n_i^-}^-\big) |0\rangle \,,
\ee
or linear combinations thereof, where $ {\cal N}_{\{n_i^\pm\}}$ is the normalization factor. 

This is one of our key results. It yields explicitly all possible BTZ microstates $|\cB(\{n_i^\pm\})\rangle$, for simplicity henceforth denoted as $|\cB\rangle$, provided our natural assumptions spelled out above hold. We emphasize that the ``horizon fluff'' \eqref{BH-microstate} form a finite subset of all soft hair descendants \eqref{generic-NH-state}. The pivotal cutoff on the soft hair spectrum is provided by the vevs $\De^\pm$ together with our relations between near horizon and asymptotic algebras.

The states $|\cB\rangle$  form an orthonormal basis for our BH microstate vector space ${\cal V}_{\cB}$, which is a subspace of the near horizon Hilbert space. The normalization constant in \eqref{BH-microstate} is fixed by requiring compatibility with \eqref{BTZ-L0},
\eq{
\langle \cB'|L_0^\pm |\cB\rangle=\Delta^\pm \delta_{\cB,\cB'}\,.
}{eq:new}

Before counting our microstates \eqref{BH-microstate} we collect additional relations that prove useful for that purpose.
The rescaling of zero mode charges \eqref{eq:LcL} implies that $|\cB\>$  describes a BH with
\be\label{cut-off}
\Delta^\pm=\<\cB|L_0^\pm|\cB\rangle \approx \frac1c \<\cB|\cL_0^\pm|\cB\rangle= \frac1c\sum_i n_i^\pm=\frac1c\ {\cal E}_{\cB}^\pm\,.
\ee
In the second equality we dropped the shift by $-1/24$ since it is irrelevant semi-classically.
We also note that the vevs \eqref{eq:new} together with \eqref{eq:J} and \eqref{eq:LJ} yield
\be
\langle \cB|J_0^{\pm\,2}|\cB\rangle= \tfrac16\,c\,\Delta^\pm\,, \qquad \langle \cB'|J_{n\neq 0}^\pm |\cB\rangle= 0\,. 
\ee

\paragraph{\textbf{Black hole microstate counting.}} Having identified all the BTZ BH microstates corresponding to a BH specified by \eqref{BTZ-microstate-def}, \eqref{eq:new}, explicitly those given by  \eqref{BH-microstate}, subject to \eqref{cut-off}, we can count them.  Physically, the derivation of the microcanonical BTZ BH entropy is reduced to the counting of different microstates \eqref{BH-microstate} with the same energy $\cal E^\pm_{\cB}$. Mathematically, this reduces to a combinatorial problem solved by Hardy and Ramanujan: the number $p(N)$ of ways a positive integer $N$ can be partitioned into non-negative integers in the limit of large $N$. The result is given by the Hardy--Ramanujan formula, see e.g.~\cite{Carlip:1998qw} and Refs.~therein:
\be\label{Hardy-Ramanujan}
p(N)\big|_{N\gg 1} \simeq \frac{1}{4N\sqrt3}\exp{\Big(2\pi\sqrt{\frac{N}{6}}\Big)}\,.
\ee
The number of BTZ BH microstates according to our discussion above yields $p({\cal E}_\cB^+)\cdot p({\cal E}_\cB^-)$.
The microcanonical entropy of a BH with $\Delta^\pm$ is then given by the logarithm of the number of microstates,
\be\label{microstate-counting}
 S = \ln p({\cal E}_{\cB}^+) + \ln p({\cal E}_{\cB}^-)= 2\pi \Big(\sqrt{\frac{c\Delta^+}{6}} +  \sqrt{\frac{c \Delta^-}{6}}\Big) + \dots
\ee
where we have assumed $c\Delta^\pm\gg 1$, and the ellipsis denotes possible corrections of order $\ln S$. 

The microscopic entropy \eqref{microstate-counting} is our second key result. We emphasize that as opposed to previous Cardy-type of microstate countings, see e.g.~\cite{Strominger:1997eq, Carlip:1998wz, Guica:2008mu, Carlip:2002be}, 
we have explicitly identified all the microstates \eqref{BH-microstate} that we counted.

Our microscopic result \eqref{microstate-counting} to leading order is exactly the Bekenstein--Hawking entropy \eqref{eq:SBH} of the BTZ BH, $S=S_{\textrm{\tiny BH}}=2\pi r_+/(4G_N)$, once we recall the relation between $\Delta^\pm$ and the inner and outer horizon radii $r_\pm$ (e.g.~see \cite{Sheikh-Jabbari:2014nya, Strominger:1997eq}),
$\Delta^\pm = \tfrac1{16\ell G_N}(r_+ \pm r_-)^2$. 

\smallskip

\paragraph{\textbf{Relations to previous approaches.}} We compare now our horizon fluff proposal with related proposals (other than AdS$_3$/CFT$_2$) and recent achievements. 

The repeated use of the notion of ``soft hair'' suggests that our results are compatible with recent work by Hawking, Perry and Strominger that introduced this notion \cite{Hawking:2016msc}. Our results also build on \cite{Afshar:2016wfy} that discovered the near horizon symmetry algebra \eqref{NH-algebra-J}. Neither of these papers attempt to identify the BH microstates. Our key insight and input here is that not all near horizon soft hairs are BH microstates, but only the horizon fluff \eqref{BH-microstate}.

Our findings are compatible with and conceptually close to Carlip's view that the BH microstates are constructed from the near horizon information \cite{Carlip:1994gy,Carlip:1998wz,Carlip:1998qw,Carlip:1999cy,Carlip:2002be}. Our proposal is, however, different as the relevant symmetry algebra \eqref{NH-algebra-J} is not a chiral Virasoro algebra.

In comparison with the fuzzball proposal \cite{Mathur:2005zp,Mathur:2014nja}, or its string theory realizations, e.g.~see \cite{Lunin:2002qf, Jejjala:2005yu, David:2002wn,  Skenderis:2008qn}, we have two distinctive features: 1.~According to our proposal BH microstates, horizon fluff, are geometries that are arranged in the representations of the near horizon algebra \eqref{NH-algebra-J} as well as the orbits of asymptotic Virasoro algebra \eqref{asymptotic-algebra-Vir}. As discussed above, the horizon fluff fall into the subset of coadjoint orbits of Virasoro algebra \eqref{asymptotic-algebra-Vir} associated with  specific conic singularities whose mass and angular momentum are integer-valued and add up to those of the BTZ BH, \emph{cf.} \eqref{cut-off}. Our horizon fluff corresponds to geometries which do not have a horizon, as in the fuzzball case.  2.~The horizon fluff are not recognizable by any observer away from the horizon, even by their asymptotic symmetry charges. In particular, to specify 
our BH microstates from a near horizon perspective we need not know about the quantum gravity theory and do not rely on extra symmetries like supersymmetry; the semiclassical description of the residual symmetry charges and associated representations are enough.


\paragraph{\textbf{Algebraic aspects of generalization to Kerr.}}
Our proposal of horizon fluff, the subset of near horizon soft hairs which cannot be distinguished away from the horizon, as BH microstates has some general features which we expect could be applicable to higher dimensional generic BHs, in particular the astrophysical BHs described by the Kerr geometry.  For example, the appearance of the Heisenberg near horizon symmetry algebra is ultimately related to 
the fact that close to the horizon the Eqs.~of motion for the modes essentially reduce to a free theory on Rindler space, which is a dimension-independent statement. 

Algebraically, a key question is whether it is again possible to construct known near horizon symmetry algebras as composites of Heisenberg algebras (or, equivalently, from $\hat u(1)$ current algebras), similar to \eqref{eq:twistedsugawara}. If it turns out that this is impossible this would constitute an algebraic {obstruction for higher dimensional generalization} of the microstate picture we are advocating in this Letter, namely as specific near horizon descendants of the vacuum. As a first step towards generalization to 4D we show now that such a construction is possible. 

We start with Heisenberg algebras or, equivalently, two $\hat u(1)$ current algebras with generators $\cJ^\pm_n$ obeying the commutation relations \eqref{NH-algebra-J}. We then intend to combine the generators non-linearly such that we recover the 4D near horizon symmetry algebra derived recently by Donnay, Giribet, Gonz\'alez and Pino \cite{Donnay:2015abr},
\begin{subequations}\label{eq:DGGP}
\begin{align}
[\cY_n^\pm,\,\cY^\pm_m] &= (n-m)\, \cY^\pm_{n+m} \\
[\cY_l^+,\,\cT_{(n,m)}] &= -n\, \cT_{(n+l,\,m)}\\
[\cY_l^-,\,\cT_{(n,m)}] &= -m\, \cT_{(n,\,m+l)}
\end{align}
\end{subequations}
where $n,m,l$ are integer values labeling Laurent modes with respect to stereographic coordinates on the horizon 2-sphere; all commutators not displayed in \eqref{eq:DGGP} vanish [the same applies to \eqref{eq:whatever2} below]. We succeed by adding a second copy of the near horizon algebra \eqref{NH-algebra-J} (that commutes with it), where for convenience we reverse the sign on the right hand side:
\be\label{eq:whatever2}
[\cJ_n^\pm,\,\cJ_m^\pm]=\tfrac12 \,n\, \delta_{n,-m}
=-[\cK_n^\pm,\,\cK_m^\pm]\,.
\ee
It is then straightforward to verify that the bi-linear combinations
\begin{subequations}
\label{eq:angelinajolie}
\begin{align}
\cT_{(n,\,m)} &= \big(\cJ_n^+ + \cK_n^+\big)\big(\cJ_m^- + \cK_m^-\big), \\ 
\cY_n^\pm &= \sum_{p\in\mathbb{Z}}  \big(\cJ^\pm_{n-p} + \cK^\pm_{n-p}\big)\big(\cJ_p^\pm - \cK^\pm_p\big),
\end{align}
\end{subequations}
generate the near horizon algebra \eqref{eq:DGGP}. Thus, there is no algebraic obstruction and the 4D near horizon symmetry algebra is indeed a composite of four $\hat u(1)$ current algebras \eqref{eq:whatever2}. This is a very remarkable result as it  indicates how our proposal of horizon fluff can be generalized to astrophysical BHs.

What remains to be done is to relate this near horizon algebra to some asymptotic one and perform a microstate counting in the spirit of the 3D calculation discussed in our present work. We hope to tackle this in future publications. 

\paragraph{Note added (October 2016).} While resubmitting this paper a number of related work has appeared \cite{ Grumiller:2016kcp, Donnay:2016ejv, Zuo:2016ezr, Cai:2016idg, Banerjee:2016nio}. In particular, see \cite{Sheikh-Jabbari:2016npa} for how horizon fluff proposal works for more general black holes with nonvanishing $J_n^\pm$ charges. 

\paragraph{Note added (May 2017).} Apart from deleting two obsolete paragraphs on the interpretation and logarithmic corrections to the entropy, the final version of this paper essentially coincides with the original one; however, due to the long period between resubmissions we made considerable advances on the interpretation of our results, a derivation of the Ba\~nados map that in the present work is introduced by fiat, and a confirmation that the logarithmic corrections to the BTZ black hole entropy are correctly accounted for by our microstates. These new results are published separately (together with Hossein Yavartanoo) \cite{Afshar:2017okz}, building essentially on the present work.

 \acknowledgements

\paragraph{Acknowledgments.} 
HA and DG thank Stephane Detournay, Wout Merbis, Blaza Oblak, Alfredo Perez, Stefan Prohazka, David Tempo and Ricardo Troncoso for collaboration on various aspects of near horizon symmetries. We thank Massimo Porrati for a crucial comment on the definition of black hole microstates given in Eqs.~\eqref{BTZ-microstate-def}, \eqref{eq:new}. We thank Finn Larsen for comments on the relation to AdS3/CFT2. We also acknowledge the scientific atmosphere of three recent conferences and workshops: \emph{Topics in three dimensional Gravity,} March 2016, the Abdus Salam
ICTP, Trieste, Italy; \emph{Recent Trends in String Theory}, IPM, Tehran, May 2016;  \emph{Recent developments in symmetries and (super)gravity theories}, IMBM,  Bogazici University, Istanbul, June 2016.
The work of HA is supported in part by Iran's National Elites Foundation 
and that of M.M. Sh-J is supported in part by Allameh Tabatabaii Prize Grant of Iran's National Elites Foundation, the SarAmadan grant of Iranian vice presidency in science and technology and the Iranian NSF junior research chair in black hole physics. M.M.Sh-J also acknowledges ICTP network project NT-04.
DG was supported by the Austrian Science Fund (FWF), projects P~27182-N27 and P~28751-N27.


%

\end{document}